\documentclass[showpacs,aps,prb,twocolumn]{revtex4}

\usepackage{amsmath}
\usepackage{amssymb}
\usepackage{amsbsy}
\usepackage{makeidx}
\usepackage{epsf}
\usepackage{graphicx}
\usepackage{amsfonts}

\begin{document}
\title{Existence and stability analysis of finite $0$-$\pi$-$0$ Josephson junctions}
\author{Saeed Ahmad}
\author{Hadi Susanto}
\author{Jonathan A.D. Wattis}
\affiliation{School of Mathematical Sciences, University of Nottingham, University Park, Nottingham, NG7 2RD, UK}


\begin{abstract}
We investigate analytically and numerically a Josephson junction on finite domain with two
$\pi$-discontinuity points characterized by a jump of $\pi$ in the
phase difference of the junction, i.e.\ a $0$-$\pi$-0 Josephson junction. The system is described by a
modified sine-Gordon equation. We show that there is an 
instability region in which semifluxons will be spontaneously
generated. Using a Hamiltonian energy characterization, it is shown how the existence of static semifluxons
depends on the length of the junction, the facet length, and the applied bias current. The critical eigenvalue of the semifluxons is discussed as well. Numerical simulations are presented, accompanying our analytical results.
\end{abstract}

\maketitle

\section{Introduction}

Josephson junctions consist of two superconductors separated by a thin insulating barrier. When the insulation is thin enough, a current can flow across the barrier even when there is no potential difference. 
Current technological advances can manipulate the flow of the supercurrent such that its direction can depend on the spatial variable. This is possible, e.g., by imposing 
 shifts to the Josephson phase.

The idea of having a 
 shift in the gauge phase of a Josephson junction was first proposed
by Bulaevskii et al.\ \cite{bula77,bula78} It was proposed that the presence of magnetic impurities may create a $\pi$-
shift to the Josephson phase, which has been confirmed recently.\cite{vavr06}
Presently, one can also impose a $\pi$ phase-shift in a long Josephson junction using superconductors with unconventional pairing symmetry,\cite{tsue00,guma07} Superconductor-Ferromagnet-Superconductor (SFS) $\pi$-junctions,\cite{ryaz01}, Superconductor-Normal metal-Superconductor (SNS) junctions,\cite{base99} or using a pair of current injectors.\cite{gold04} All these findings have promising applications in information storage and information processing.\cite{hilg03}

This system, 
 in which neighboring facets 
   of a Josephson junction can be considered to have opposite 
    signs of the critical current, present intriguing phenomena such as the intrinsic frustration of the Josephson phase over the junction and the spontaneous generation of a fractional magnetic flux at the 
     discontinuities i.e.\ the position of the jump in the Josephson phase.\cite{hilg03,frol08} 

In the present work, we consider the so-called 0-$\pi$-0 Josephson junctions on a finite domain, modeled by a modified sine-Gordon equation with phase shift of $\theta=\pi$ in some region and zero otherwise. 
An infinite domain 
 0-$\pi$-0 Josephson junction was first studied by Kato and Imada,\cite{ki} where they showed that there is a 
 stability window for the $\pi$ junction length in which the zero constant  solution is 
 stable. In the instability region, the ground state is 
   non constant solution in space, which corresponds to antiferromagnetically ordered semifluxons. Later, using an existence analysis, it was shown that the presence of 
   a minimum facet length of the $\pi$ junction, above which a non-trivial ground state emerges, also corresponds to the minimum facet length needed to construct such solutions.\cite{zenc04,susa03} The possibility 
      of employing 0-$\pi$-0 junctions for observations of 
      macroscopic quantum tunneling was discussed in length by Goldobin et al.\cite{gold05_1} In the presence of an applied bias current, a 0-$\pi$-0 Josephson junction has a critical current above which one can thermally flip the order of the semifluxons\cite{ki} and another critical current above which the junction switches to the resistive state.\cite{susa03} Goldobin et al.\cite{gold04_2,gold05_2} have also 
      broadened the study of 0-$\pi$-0 junctions to 0-$\kappa$-0 junctions, where $0\leq\kappa\leq\pi$ (mod $2\pi$). Here, we limit ourselves to discuss 0-$\pi$-0 junctions only, but extend it to the case of a finite domain. This is of particular interest, especially from the physical point of view, as such junctions have been successfully fabricated recently,\cite{dewe08,hans06} which are certainly of finite length.

The present 
paper is structured as follows. In Section II, we discuss the mathematical model that we use to describe the problem. We then show in Section III that when there is no bias current, the equation has two constant solutions. Due to the phase-shifts, there will be a region of facet lengths, in which both constant solutions are unstable. In 
this instability region, a non-constant ground state will emerge from the constant solutions, i.e.\ a pair of semifluxons is the ground state of the system. A Hamitonian analysis is performed in Section IV to study the behavior of the non-trivial ground state, both with and without the presence of applied bias current. We then compare our analytical results with numerical computations in Section V. Finally, conclusions are presented in Section VI.

\section{Mathematical model}
\label{sec2}

The dynamics of a finite Josephson junction with
$\pi$-discontinuity points is commonly described by the following perturbed sine-Gordon equation
\begin{equation}
\phi_{xx}-\phi_{tt}=\sin(\phi+\theta(x))-\gamma+\alpha\phi_t, \quad -L \leq x \leq L,
\label{sine-gordon}
\end{equation}
where $\alpha$ is a dimensionless positive damping coefficient
related to quasi-particle tunneling across the junction, $L$ is the length of the junction, and
$\gamma$ is the applied bias current density normalized to the
junction critical current density $J_c$. Without loss of generality, 
in the following we set  $\alpha=0$.

Equation (\ref{sine-gordon}) is written after rescaling where the
spatial variable $x$ and time variable $t$ are normalized to the
Josephson penetration length $\lambda_J$ and the inverse plasma
frequency $\omega_p^{-1}$, respectively.

The function $\theta$, representing the presence, or absence, of the
additional $\pi$-phase shift, is given by
     \begin{equation}
\theta(x)=\left\{\begin{array}{cc}
0, & L>|x|>a, \\
\pi, & |x|<a,\\
\end{array}\right. \label{saud2}
\end{equation}
where $a$  is the length of the $\pi$ junction, i.e.\ the facet length.

Equation (\ref{sine-gordon}) is subject to the continuity and boundary conditions
    \begin{equation}
     \phi(\pm a^-)=\phi(\pm a^+),\,\phi_x(\pm a^-)=\phi_x(\pm a^+),\,\phi_{x}(\pm L)=0.
     \label{saud3}\end{equation}

The governing equation (\ref{sine-gordon}), subject to the boundary conditions (\ref{saud3}), can be derived from the Lagrangian
\begin{equation}
\mathcal{L}=\int_{-L}^{L}(\frac12\phi_t^2-\frac{1}{2}\phi^2_{x}-1+\cos(\phi+\theta)+\gamma\phi)\,dx.
\label{hamilton}
\end{equation}

As we mainly consider static semifluxons, the existence of the solutions will be studied through the time-independent 
version of (\ref{sine-gordon}), namely
\begin{equation}
\phi_{xx}=\sin(\phi+\theta)-\gamma. \label{sgstatic}
\end{equation}

\section{Existence and stability analysis of constant solutions}
	
It is clear that  equation\ (\ref{sgstatic}) admits two constant solutions (modulo $2\pi$), namely
\[\tilde\phi=\arcsin\gamma,\,\pi-\arcsin\gamma,\]
for $a<|x|<L$, and
\[\tilde\phi=\arcsin\gamma-\pi,\,-\arcsin\gamma,\]
for $0<|x|<a$.

As solutions on the whole domain must satisfy the continuity conditions (\ref{saud3}), it can be concluded that constant solutions exist only when $\gamma=0$, i.e.\ $\tilde\phi=0$ and $\tilde\phi=\pi$.

Next, we will determine the linear stability of the constant solutions. For this purpose, we substitute the stability ansatz
\begin{equation}
\phi=\tilde\phi+\epsilon e^{\lambda t}V(x)
\label{stab}
\end{equation}
into Eq.\ (\ref{sine-gordon}). Neglecting higher order terms, one 
obtains the 
 eigenvalue problem
\begin{equation}
V_{xx}=(E+\cos(\tilde\phi+\theta))V,
\label{EVP}
\end{equation}
where $E=\lambda^2$ and $V$ is also subject to the continuity and boundary conditions
    \begin{equation}
     V(\pm a^-)=V(\pm a^+),\,V_x(\pm a^-)=V_x(\pm a^+),\,V_{x}(\pm L)=0.
     \label{cc}
     \end{equation}

From the stability ansatz (\ref{stab}), it is clear that $\tilde\phi$ is stable if $\lambda$ has zero real parts. Therefore, a solution $\tilde\phi$ is said to be linearly stable if $E<0$ and unstable when $E>0$.

Due to the finite size of the domain, the eigenvalue problem (\ref{EVP}) will give two sets of eigenvalues, i.e.\ infinitely many eigenvalues that will constitute the continuous spectrum in the limit $L\to\infty$ and a finite number 
in the discrete spectrum. For simplicity, 
in the following sections we denote the infinitely many eigenvalues by 
the 'continuous' spectrum.

\subsection {Linear stability of $0$}

First, we 
discuss the 'continuous' spectrum of $\tilde\phi=0$.

Looking for a bounded solution to (\ref{EVP}) that satisfies the boundary conditions at $x=\pm L$, we obtain the 
solutions
\begin{equation}
V(x)=\left\{
\begin{array}{llll}
&&A\cos(\hat{\alpha}(x+L)),\,-L<x<-a,\\
&&B\cos(\hat{\beta}x)+C\sin(\hat{\beta}x),\,|x|<a,\\
&&D\cos(\hat{\alpha}(x-L)),\,a<x<L,
\end{array}
\right.
\end{equation}
where $\hat{\alpha}=\sqrt{-1-E}$ and $\hat{\beta}=\sqrt{1-E}$.

Using the continuity conditions (\ref{cc}), we 
 obtain a system of four equations with four unknowns, given in a matrix form by
\[M_1
\left(
\begin{array}{c}
A\\B\\C\\D
\end{array}
\right)=0,
\]
with a coefficient matrix $M_1$ given in the appendix.

\begin{figure}
\begin{center}
\includegraphics[width=7cm]{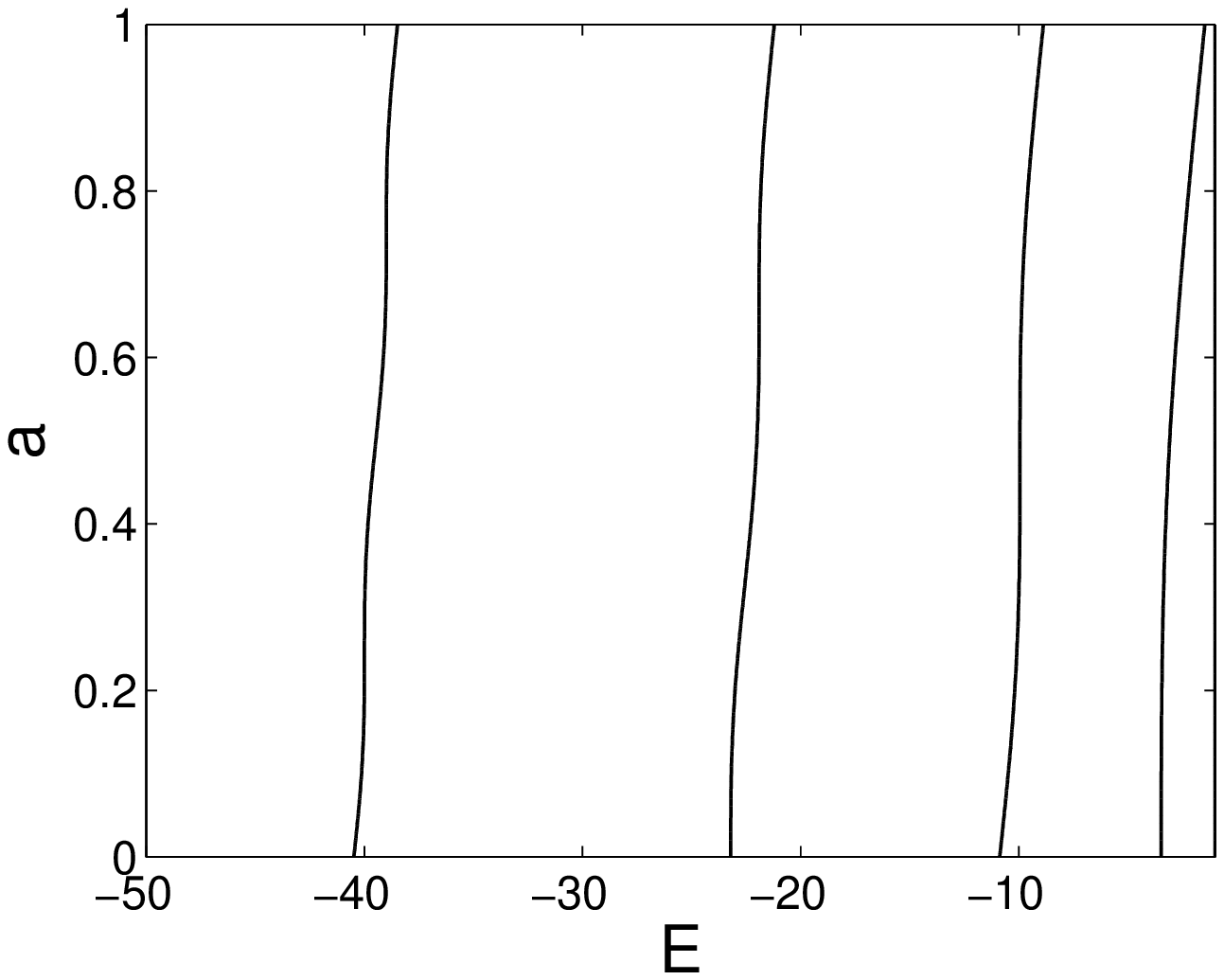}\\
\includegraphics[width=7cm]{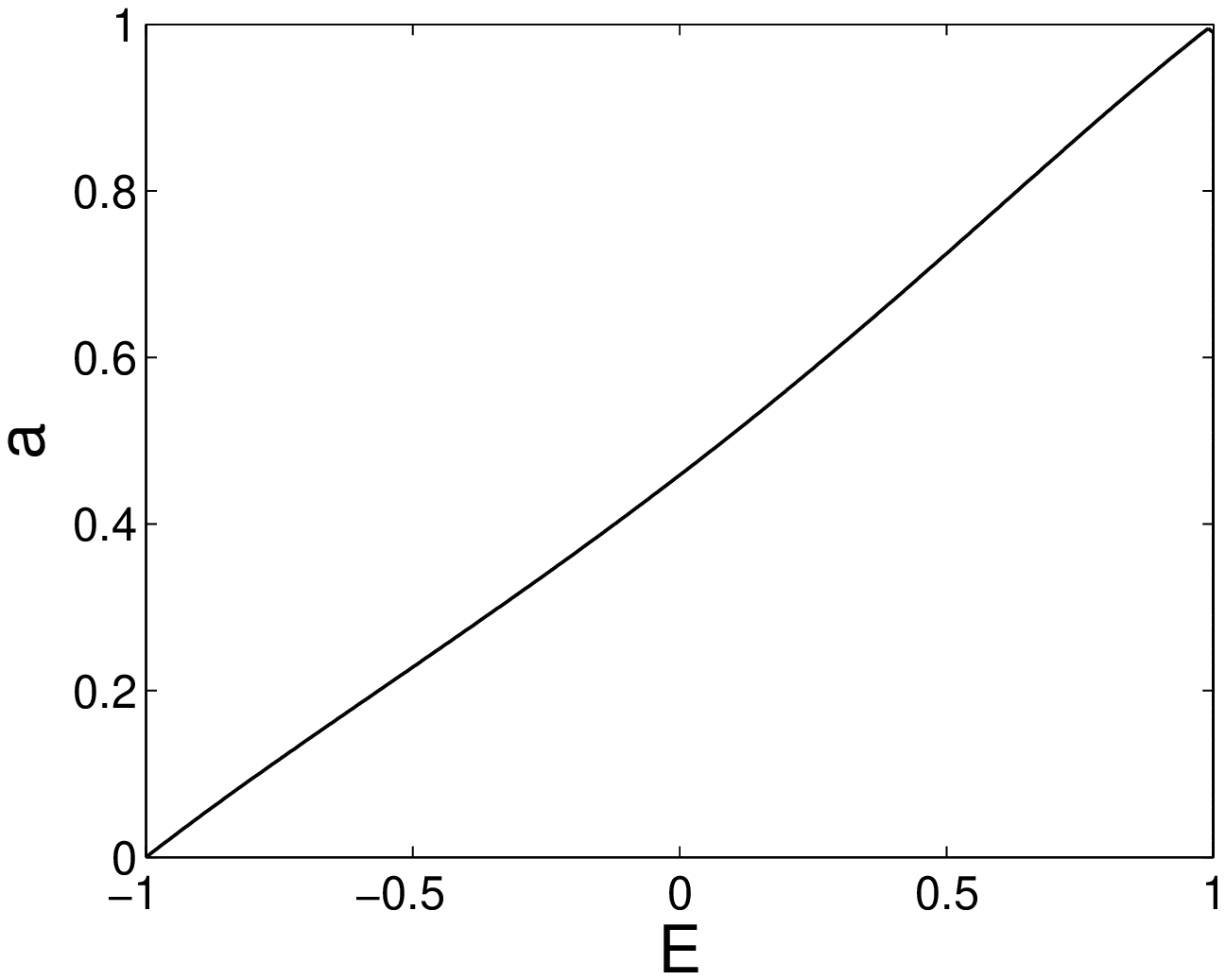}
\end{center}
\caption{Plot of the 'continuous' spectrum (top) and the discrete spectrum (bottom) of $\tilde\phi=0$ as a function of the $\pi$ junction length for $L=1.$ }
\label{contspectofzero}
\end{figure}

To obtain a nontrivial $V$, 
 we require det$(M_1)=0$. An implicit plot of the equation, i.e.\ the 'continuous' spectrum $E(a,L)$, for $L=1$ is shown in the top panel of Fig.\ \ref{contspectofzero}.  Numerically, it is found that there is no unstable eigenvalue in the 'continuous' spectrum, i.e.\ $E<0$ for any $a$. As $L$ increases, the distribution of $E$ will become dense, as expected.

Next, we  
find the discrete spectrum of the constant solution $\tilde\phi=0$, corresponding to bounded and decaying solutions 
 of the eigenvalue problem (\ref{EVP}). We 
 obtain 
 the solution 
\begin{equation}
V(x)=\left\{
\begin{array}{llll}
&&A\cosh(\hat{\gamma}(x+L)),\,-L<x<-a,\\
&&B\cos(\hat{\beta}x)+C\sin(\hat{\beta}x),\,|x|<a,\\
&&D\cosh(\hat{\gamma}(x-L)),\,a<x<L,
\end{array}
\right.
\label{V0}
\end{equation}
where $\hat{\gamma}=\sqrt{1+E}$  and $\hat{\beta}$ is defined above.

From the continuity conditions, again we 
 find a system of four equations. As 
  above, the eigenvalues are obtained 
  by setting the determinant of 
  the coefficient matrix $M_2$, given in the appendix, to zero. An implicit plot of the eigenvalues as a function of $a$ for $L=1$ is shown in the bottom panel of Fig.\ \ref{contspectofzero}.

From Fig.\ \ref{contspectofzero}, we 
observe that for a given $L$, there is a critical $a$ above which $E$ becomes positive, i.e., 
 $\tilde{\phi}=0$  becomes unstable. In the following, we 
  denote such a critical $a$ by $a_{c,0}$. For $L=1$, $a_{c,0}\approx0.46$. As $L$ increases, $a_{c,0}$ will asymptotically approach $\frac{\pi}{4}$, which is the critical facet length in the infinite domain calculated in Refs.\ \onlinecite{ki,susa03}. The relation between $a_{c,0}$ and $L$ is implicitly given by the smallest positive root of
\begin{equation}
\cot(a_{c,0})\tanh(L-a_{c,0})=-1,
\label{ac0}
\end{equation}
which is obtained by considering the even mode of (\ref{V0}), i.e.\ setting 
$E=C=0$ and $A=D=B\cos(a_{c,0})/\cosh(L-a_{c,0})$. For 
small  $L$ the root can be approximated by
\begin{equation}
a_{c,0}=\frac{L}{2}-\frac{1}{24}L^3+\mathcal{O}(L^5).
\label{ac0appr}
\end{equation}

Plots 
of $a_{c,0}$ as a function of $L$ given implicitly by (\ref{ac0}) and its approximation (\ref{ac0appr}) are shown in Fig.\ \ref{combinedinstability}.

\subsection{Linear stability of $\pi$}

Following the same steps as we did in the stability analysis of 
$\tilde{\phi}=0$, the solution $V$ to the eigenvalue problem (\ref{EVP}) that corresponds to the 'continuous' spectrum is given by
\begin{equation}
V(x)=\left\{
\begin{array}{llll}
&&A\cos(\hat{\beta}(x+L)),\,-L<x<-a,\\
&&B\cos(\hat{\alpha}x)+C\sin(\hat{\alpha}x),\,|x|<a,\\
&&D\cos(\hat{\beta}(x-L)),\,a<x<L.
\end{array}
\right.
\end{equation}
One 
then finds 
that the spectrum is given by the zero of the determinant of  
the coefficient matrix $M_3$, given in the appendix.

\begin{figure}
\begin{center}
\includegraphics[width=7cm]{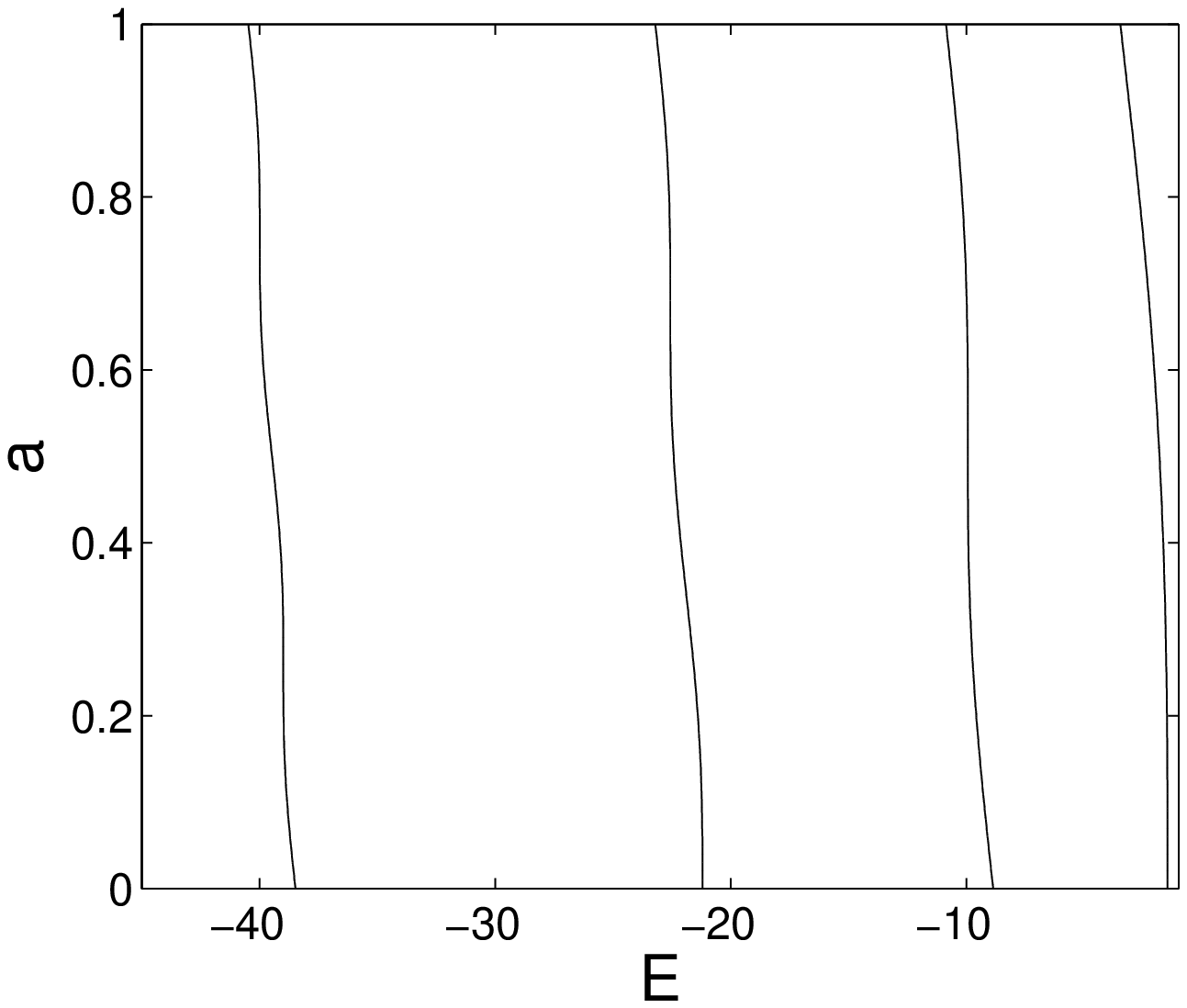}\\
\includegraphics[width=7cm]{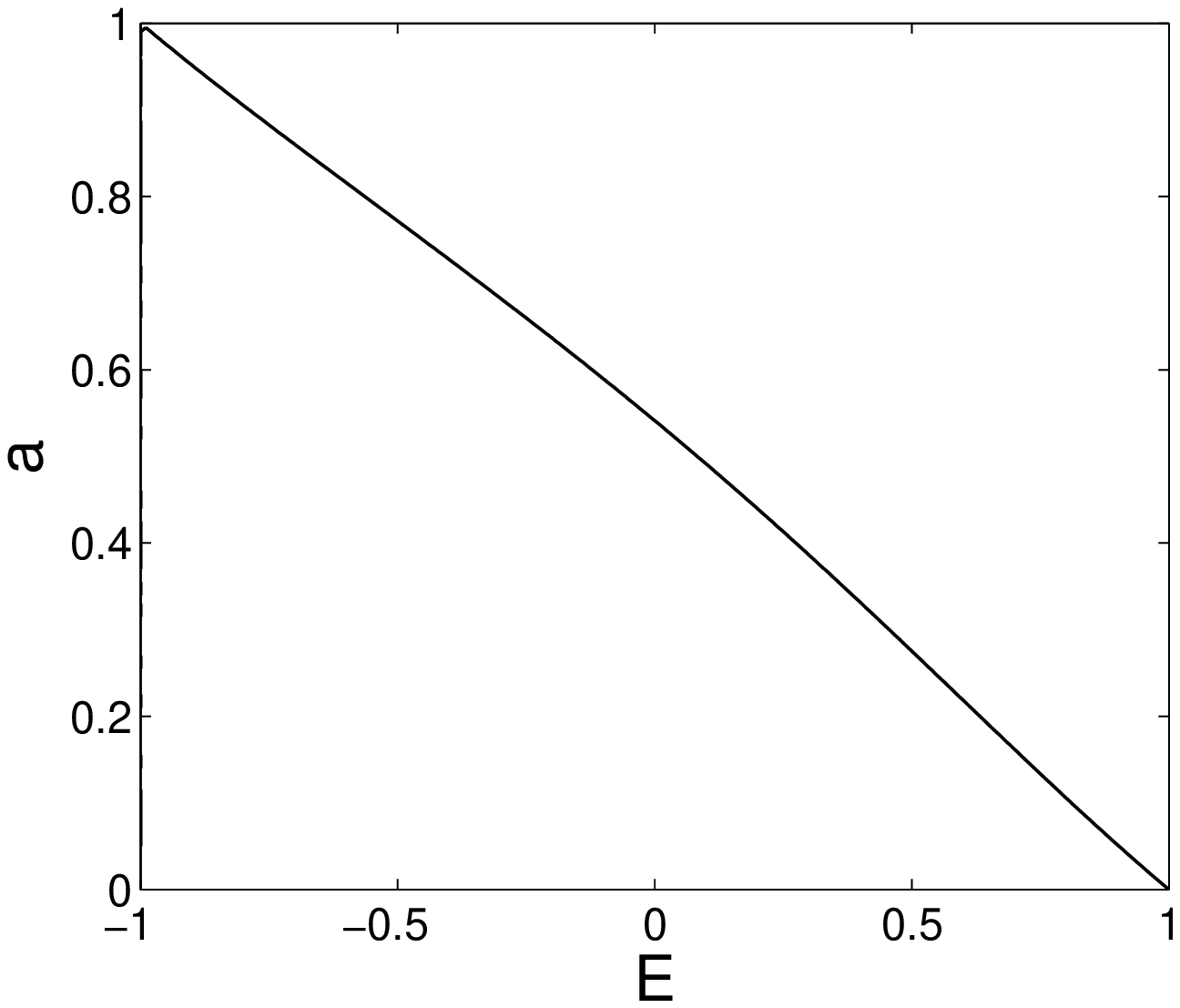}
\end{center}
\caption{The same as Fig.\ \ref{contspectofzero}, but for $\tilde\phi=\pi$.}
\label{Contspectrumofpi}
\end{figure}

A plot of the 'continuous' spectrum in the ($E,a$)-plane is shown in Fig.\ \ref{Contspectrumofpi}, from which it is clear that the continuous spectrum also only consists of stable eigenvalues.

For the discrete spectrum of 
$\tilde{\phi}=\pi$ in a finite domain, bounded and decaying solutions $V$ to the eigenvalue problem (\ref{EVP}) are given by
\begin{equation}
V(x)=\left\{
\begin{array}{llll}
&&A\cos(\hat{\beta}(x+L)),\,-L<x<-a,\\
&&B\cosh(\hat{\gamma}x)+C\sinh(\hat{\gamma}x),\,|x|<a,\\
&&D\cos(\hat{\beta}(x-L)),\,a<x<L.
\end{array}
\right.
\label{Vpi}
\end{equation}

Due to the boundary conditions (\ref{cc}),  again we obtain a system of four homogenous equations with a coefficient matrix $M_4$, given in the appendix.

The bottom panel of Figure \ref{Contspectrumofpi} shows the plot of the zeros of det$(M_4)$ in the ($E,a$)-plane, for $L=1$. 
We observe that for $a$ close to zero, $E>0$, i.e.\ 
 $\tilde{\phi}=\pi$ is unstable. Yet, there is a critical value of $a$ above which $\pi$ is stable. We denote this critical facet length by $a_{c,\pi}$, which for $L=1$ is approximately $0.54$.

Again, considering the even state of (\ref{Vpi}), i.e.\ $E=C=0$ and $A=D=B\cosh(a_{c,\pi})/\cos(L-a_{c,\pi})$, one can show that $a_{c,\pi}$ is related to $L$ by the implicit equation
\begin{equation}
\coth(a_{c,\pi})\tan(L-a_{c,\pi})=1,
\label{acpi}
\end{equation}
which, for $L$ close to 0, can be approximated by
\begin{equation}
a_{c,\pi}=\frac{L}{2}+\frac{1}{24}L^3+\mathcal{O}(L^5).
\label{acpiappr}
\end{equation}

\begin{figure}
\begin{center}
\includegraphics[width=7cm]{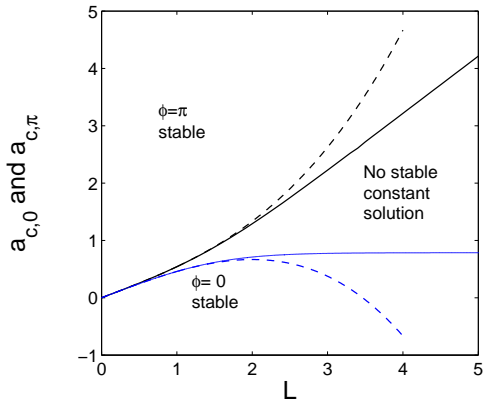}
\end{center}
\caption{Instability region of the constant solutions $\tilde\phi=0$ and $\tilde\phi=\pi$. Solid boundary curves are given by Eqs.\ (\ref{ac0}) and (\ref{acpi}). Dashed lines are analytical approximations given by (\ref{ac0appr}) and (\ref{acpiappr}).}
\label{combinedinstability}
\end{figure}

Combining Eqs.\ (\ref{ac0}) and (\ref{acpi}), we plot in Fig.\ \ref{combinedinstability} the region in which {\em both}  the stationary solutions of the sine-Gordon  (\ref{sine-gordon}) are unstable. In the instability region, the ground state will be 
non constant 
 in space.

\subsection{Symmetry}
\label{symmetry}

Comparing Figs.\ \ref{contspectofzero} and \ref{Contspectrumofpi}, 
 we observe that they are the same by reflection with respect to the line $a=L/2$, i.e.\ the stability of $\phi=0$ at the $\pi$ junction length $a$ is the same as the stability of $\phi=\pi$ at the $\pi$ facet length $(L-a)$. This symmetry occurs because, for the particular solutions, our Neumann boundary conditions at $x=\pm L$ (\ref{saud3}) can be replaced by periodic boundary conditions
\begin{equation}
\phi(-L)=\phi(L),\quad\phi_x(-L)=\phi_x(L).
\label{pbc}
\end{equation}
For 
the periodic boundary conditions, the governing equation (\ref{sine-gordon}) is symmetric by rotation, i.e.\ cyclic symmetry, and $\theta\to\theta+\pi$. Using the symmetry, one can also conclude that
\[a_{c,\pi}=L-a_{c,0},\]
for any $L$.

\section{Ground states in the instability region}

In the following, we 
 analyse perturbatively the ground states of the Josephson junction in the instability region. Our analysis, based on an Euler-Lagrange approximation, 
 is carried out for 
 a facet length $a$ close to one of the critical facet lengths $a_{c,0}$ and $a_{c,\pi}$. 

\subsection{The case of $0<a-a_{c,0}\ll 1$}

\subsubsection{Existence analysis}

For $a$ close to $a_{c,0}$, we approximate $\phi(x)$ by
\begin{equation}
\phi(x)=
B\left\{
\begin{array}{lll}
\displaystyle\frac{\cos(a_{c,0})}{\cosh(L-a_{c,0})}\cosh(x+L),\,a<|x|<L,\\
\cos(x),\,|x|<a,
\end{array}
\right.
\label{app0}
\end{equation}
where $B=B(t)$ and $a_{c,0}$ is given in (\ref{ac0}). This expression of $\phi$ is an exact solution to the 
 linearization of (\ref{sgstatic}) for $a=a_{c,0}$ and an arbitrary parameter $B$, i.e.\ we approximate the ground states by $\phi=V(x)$ , where $V(x)$ is the first even state to the eigenvalue problem (\ref{EVP}) at $E=0$ and $a=a_{c,0}$. 

Substituting the ansatz (\ref{app0}) into the Lagrangian (\ref{hamilton}), writing $a=L/2-kL^3\,k<1/24$, and expanding about $L=0$ 
 yields
\begin{equation}
\mathcal{L}=(L-L^3/4)B_t^2-H,
\label{L}
\end{equation}
where the subscript represents a derivative and
\begin{eqnarray}
H&=&\frac{L^3B^2}{12}\left(24k-1-\frac{B^2}{3}(6k-1)\right)\nonumber\\
&&+2L\left(L^2(\frac{B\gamma}{8}-4k)+1-B\gamma \right).
\label{H}
\end{eqnarray}

The Euler-Lagrange equation from the Lagrangian (\ref{L}) is $\partial_t(\partial_{B_t}\mathcal{L})-\partial_B\mathcal{L}=0,$ 
giving
\begin{equation}
B_{tt}=\frac1{2L-L^3/2}H_B.
\label{EL}
\end{equation}

The time independent solution $B=B_0$ of the Euler-Lagrange equation (\ref{EL}) is given by a cubic equation $H_B=0$, or \ 
\begin{equation}
\gamma(B_0)=-\frac{2L^2B_0(B_0^2(12k-2)+3-72k)}{9(L^2-8)}.
\label{gm}
\end{equation}
For a general value of $\gamma\neq0$, 
we solve the cubic equation using Nickalls' method\cite{nick93} to obtain
\begin{equation}
B_0^{(n)}=2\Delta\cos(\Theta+2(n-1)\pi/3),\quad n=1,2,3,
\label{B0}
\end{equation}
where
\[\begin{array}{ccc}
\Delta=\sqrt{\frac{1-24k}{2-12k}},\,\Theta=\arccos{(-y_N/h)}/3,\\
y_N=(2L-L^3/4)\gamma,\,h=-\frac1{9}L^3(1-24k)\Delta. 
\end{array}\]
When $\gamma=0$, the expressions 
for $B_0^{(n)}$ are simplified to
\begin{equation}
B_0^{(1,2)}=\pm\sqrt{\frac{3(24k-1)}{2(6k-1)}},\,B_0^{(3)}=0.
\label{s0}
\end{equation}
The non-zero roots $B_0^{(1,2)}$ represent a pair of antiferromagnetically ordered semifluxons.

If 
we study further, the three roots (\ref{B0}), we find that they do not persist for 
all $\gamma$. If $\gamma$ is decreased (increased) away from zero, then there 
is a critical value of the bias current at which $B_0^{(1)}$ ($B_0^{(2)}$) collides with $B_0^{(3)}$ in a saddle node bifurcation. Here, we denote 
this critical value of $\gamma$ by $\gamma_{c,1}$. 
From 
our current approximation, 
$\gamma_{c,1} $can be calculated from the condition $y_N^2=h^2$,\cite{nick93} which gives
\begin{equation}
\gamma_{c,1}=\frac{2\sqrt2L^2(24k-1)^{3/2}}{9\sqrt{6k-1}(L^2-8)}.
\label{gcappr}
\end{equation}

\subsubsection{Stability analysis}

To study the stability of the stationary solutions (\ref{B0}), 
we easily check that when $k<1/24$, $H$ is locally 
 minimized by  $B_0^{(1,2)}$. To obtain the critical eigenvalue of the stable solutions, we write $B=B_0^{(n)}+\epsilon\widetilde{B}$ and substitute it into the Euler-Lagrange equation (\ref{EL}) to obtain
\begin{equation}
\ddot{\widetilde{B}}(t)=\frac1{2L-L^3/2}\left.\partial_{B}^2H\right|_{B=B_0^{(n)}}\widetilde{B}.
\end{equation}
The critical eigenvalue of $B_0^{(n)}$ is then given by
\begin{equation}
E=\frac1{(2L-L^3/2)}\left.\partial_{B}^2H\right|_{B=B_0^{(n)}},
\label{EB0}
\end{equation}
i.e.\ the square of the oscillation frequency of $\widetilde{B}(t)$.

\subsection{The case of $0<a_{c,\pi}-a\ll1$}

To discuss the existence and 
stability of 
ground state solutions 
when $a$ is close to $a_{c,\pi}$, 
we repeat the 
 above calculations. We 
 exploit the symmetry discussed in \ref{symmetry}, 
 that for the non-constant ground state, 
 the Neumann boundary conditions (\ref{saud3}) can be replaced by 
 periodic boundary conditions (\ref{pbc}). Using 
 this symmetry, we obtain that if $\phi(x;a)$ is a ground state solution of the sine-Gordon equation with the $\pi$ facet length $a$, then
\begin{equation}
\phi(x;L-a)=
\pi-\phi(L-|x|;a).
\end{equation}
Thus, 
the stability of the ground state in the limit $0<a_{c,\pi}-a\ll1$ can 
be deduced using 
this symmetry argument.

\begin{figure}[tbhp]
\begin{center}
\includegraphics[width=7cm]{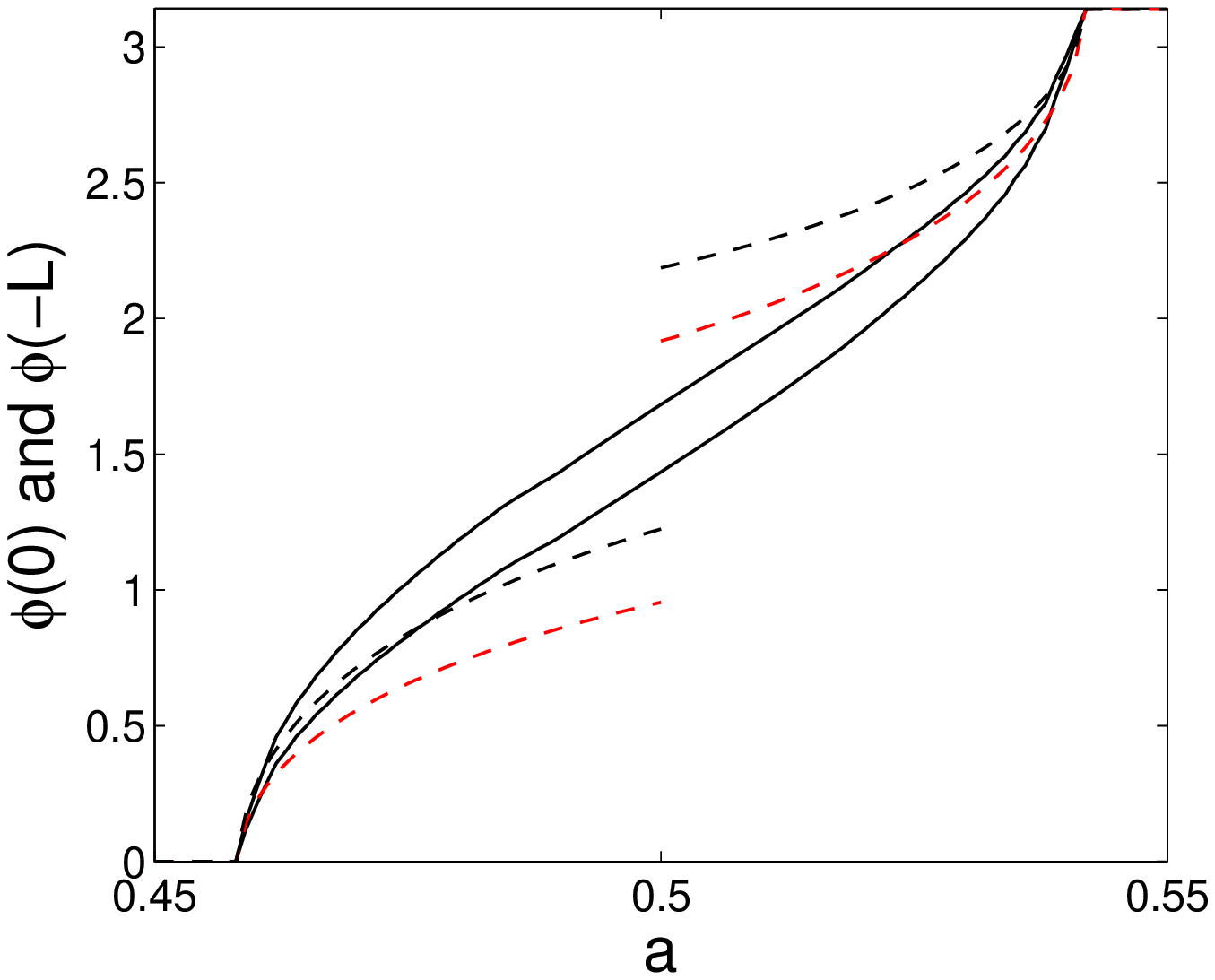}\\
\includegraphics[width=7cm]{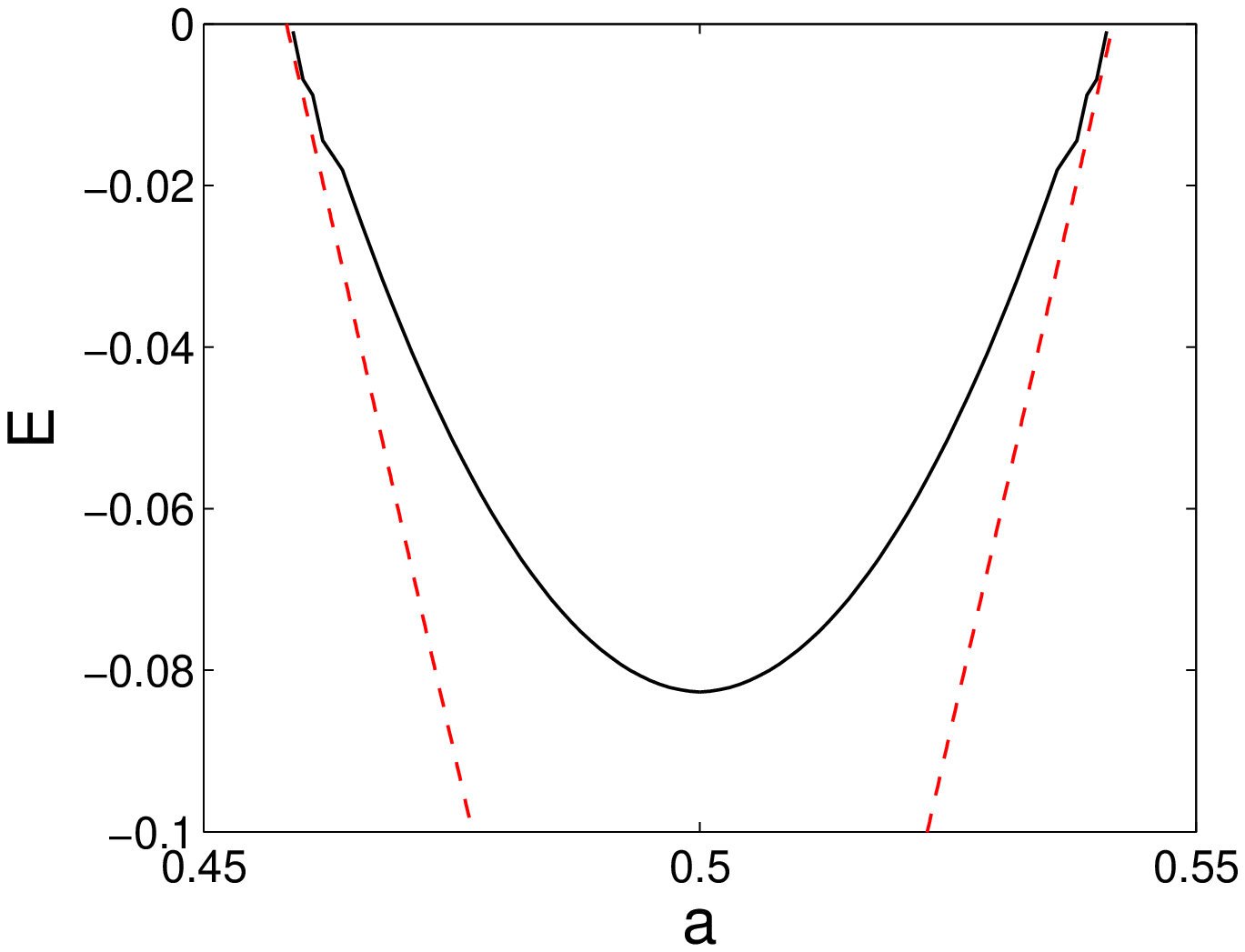}
\end{center}
\caption{(Top) Plot of $\phi(0)$ and $\phi(\pm L)$ of the non-constant ground state obtained from numerical calculations (solid lines) as a function of the facet length $a$. Comparison with our analytical approximations (dashed lines) is also presented. (Bottom) The critical eigenvalue $E$ of the solution depicted in the top panel.}
\label{approximations}
\end{figure}

\section{Discussion}

To check our analytical results, we perform numerical calculations and simulations. We 
numerically solve the time-independent governing equation (\ref{sgstatic}), subject to boundary conditions (\ref{saud3}) 
 using a Newton-Raphson method, where we 
 discretize the problem using  central 
 differences with a relatively fine spatial discretization. To numerically study the stability of 
 a solution, we then discretize and solve 
 the corresponding linear eigenvalue problem (cf.\ Eq.\ (\ref{EVP})).

First, we 
study the existence and the stability of the 
non zero ground state in the absence of an applied bias current.

In 
 the top panel of Fig.\ \ref{approximations}, we plot $\phi(x)$, which is represented by $\phi(0)$ and $\phi(\pm L)$, of the non-constant ground states as functions of $a$ for $\gamma=0$ and $L=1$. Due to the cyclic symmetry discussed in Section \ref{symmetry}, 
  we observe that the curves are symmetric under rotation by $\pi$ 
  radians. In the same figure, we also depict our 
  analytic approximation (\ref{s0}), where one can see a rather good agreement for $a$ near 
  to $a_{c,0}$ and $a_{c,\pi}$. The approximations deviate rapidly as $a$ moves away from 
  these critical lengths due to the fact that our junction length $L$ is of order one. It is expected that for $L$ close to 
  zero, the approximations will give a rather good agreement.

In 
the bottom panel of the same figure, we depict the critical eigenvalue of the non-zero ground states presented at the top panel. At the critical facet lengths, the eigenvalues are certainly zero due to the change of stability with the constant solutions $\phi=0,\,\pi$. Our analytical approximation (\ref{EB0}) is 
plotted 
in the same figure, from which 
we see that 
when the facet length $a$ close to one of the critical values $a_{c,0}$ and $a_{c,\pi}$, the numerics are indeed well approximated by our analytical result.

\begin{figure}[tbhp]
\begin{center}
\includegraphics[width=7cm]{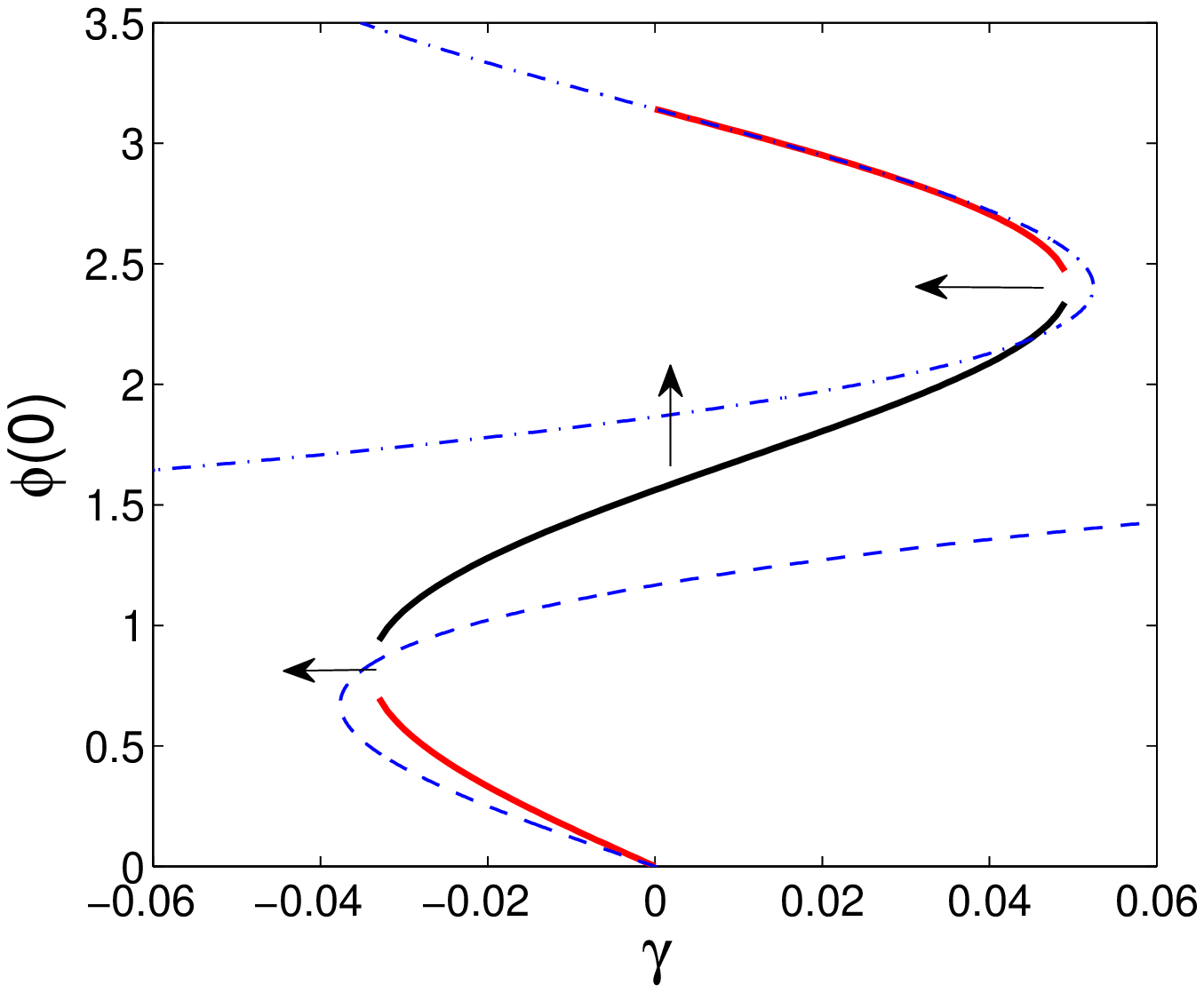}\\
\includegraphics[width=7cm]{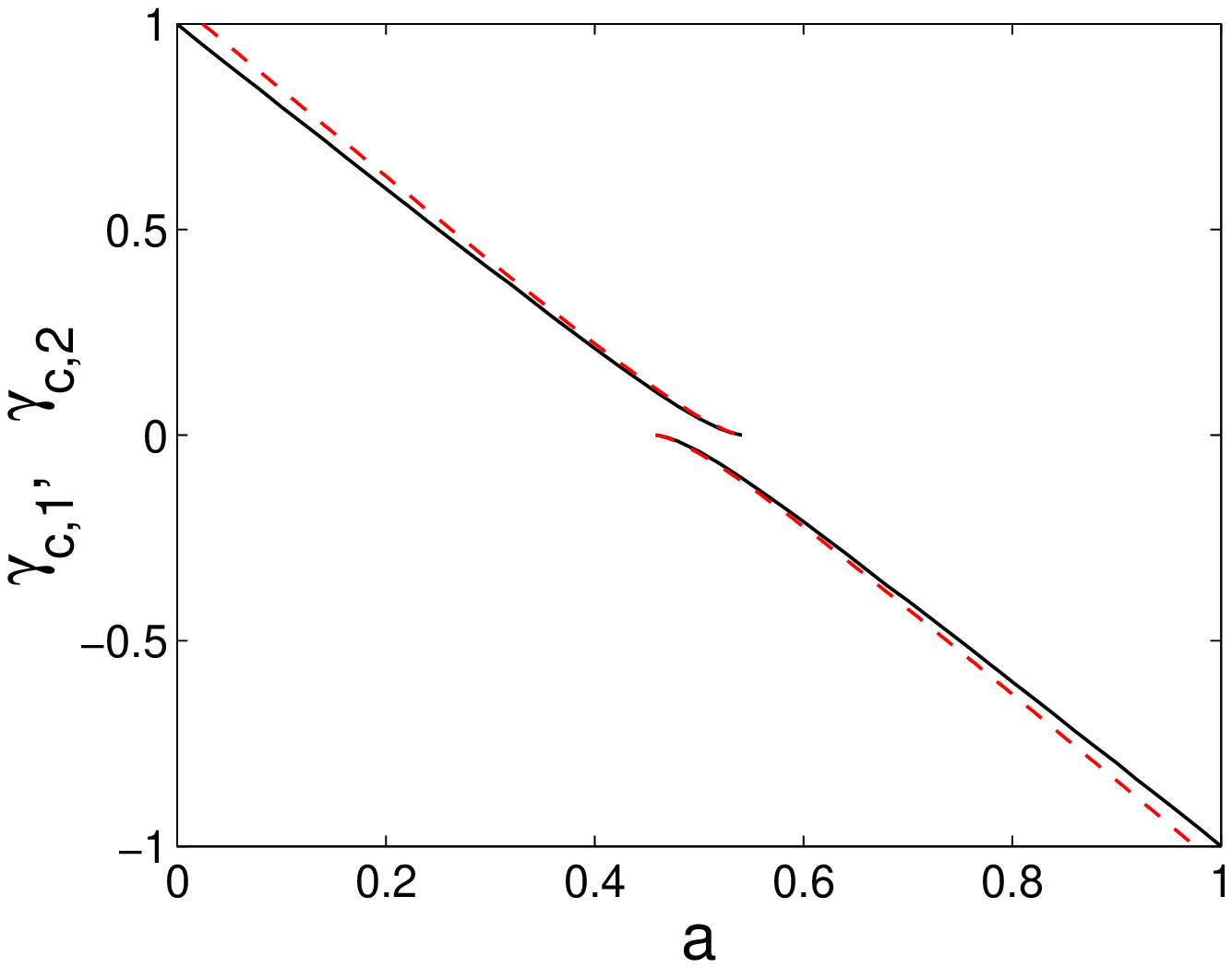}\\
\includegraphics[width=7cm]{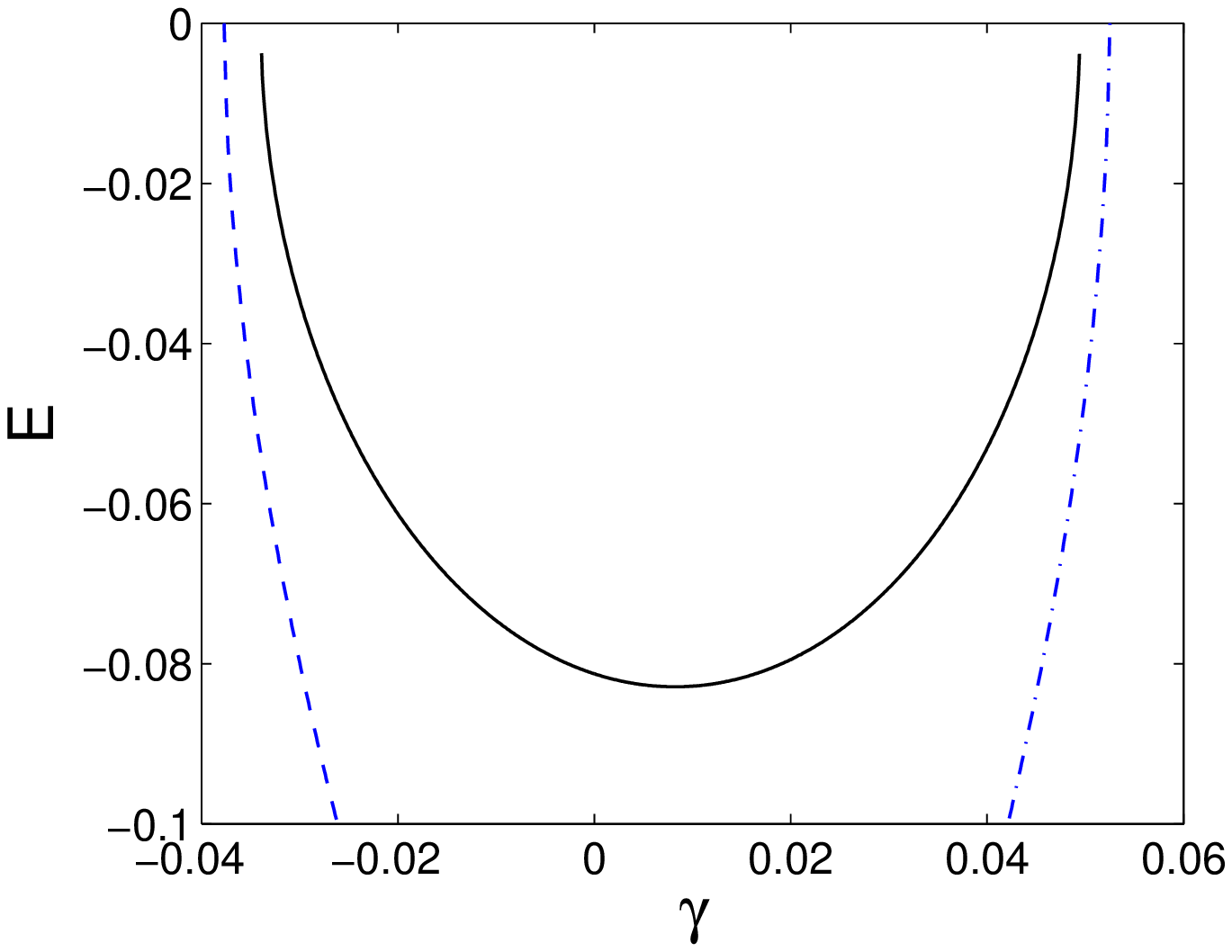}
\end{center}
\caption{(Color online) The top panel depicts the existence diagram of the ground state. Plotted is $\phi(0)$ as a function of $\gamma$, obtained from numerical computations (solid lines) for $a=0.495$ and $L=1$. Shown in red is $\phi(0)$ as a function of $a$ that corresponds unstable solutions. The upper and lower red branch corresponds to solutions $\tilde\phi=\pi,\,0$, respectively. The middle panel shows the critical bias currents $\gamma_{c,1}$ and $\gamma_{c,2}$ as a function of $a$ for $L=1$. The bottom panel presents the critical eigenvalue of the non-constant ground state as a function of $\gamma$ for $a=0.495$ and $L=1$. Analytical approximations are also presented in dashed and dash-dotted lines.}
\label{gnonzero}
\end{figure}

\begin{figure}[tbhp]
\begin{center}
\includegraphics[width=7cm]{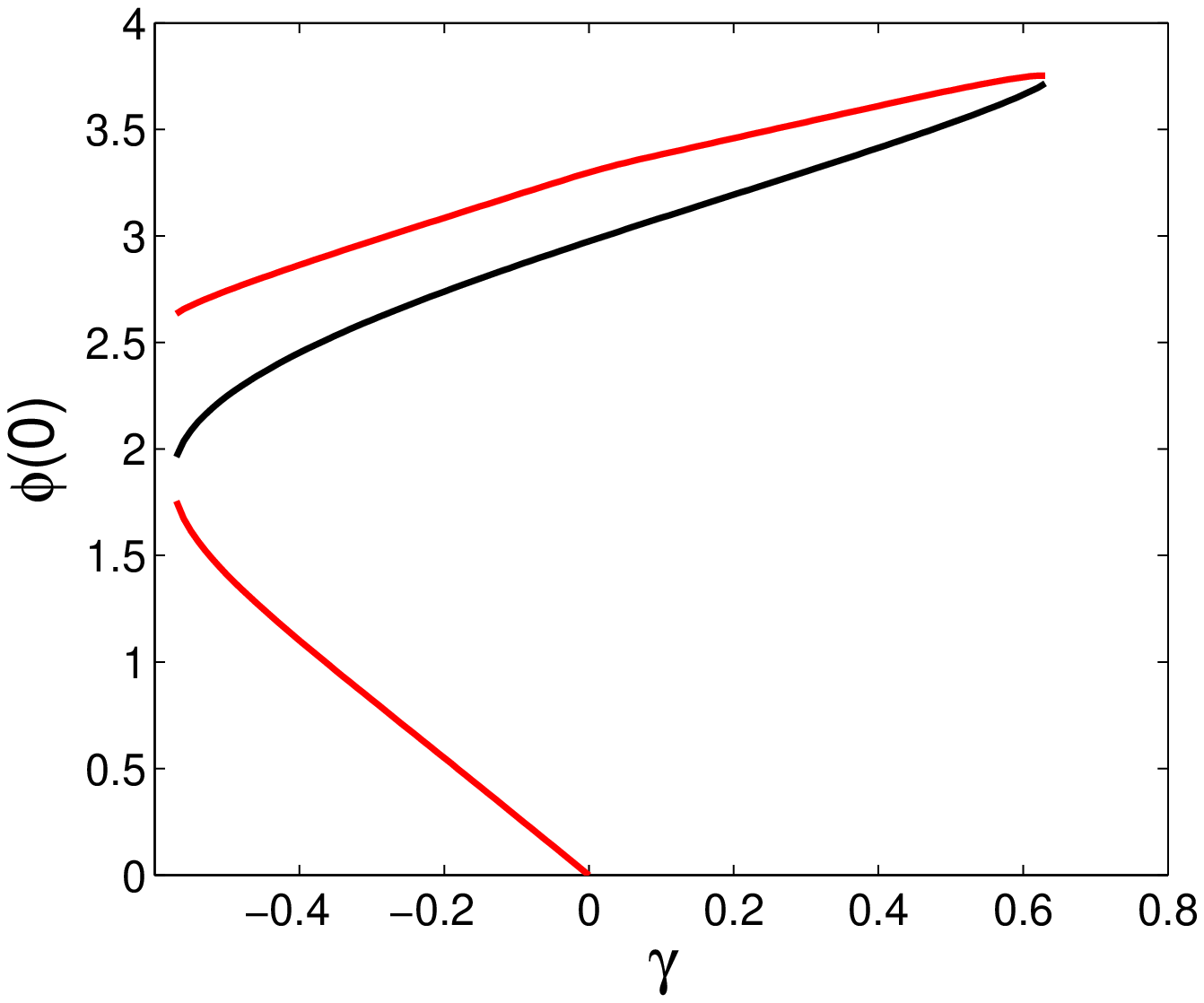}\\
\includegraphics[width=7cm]{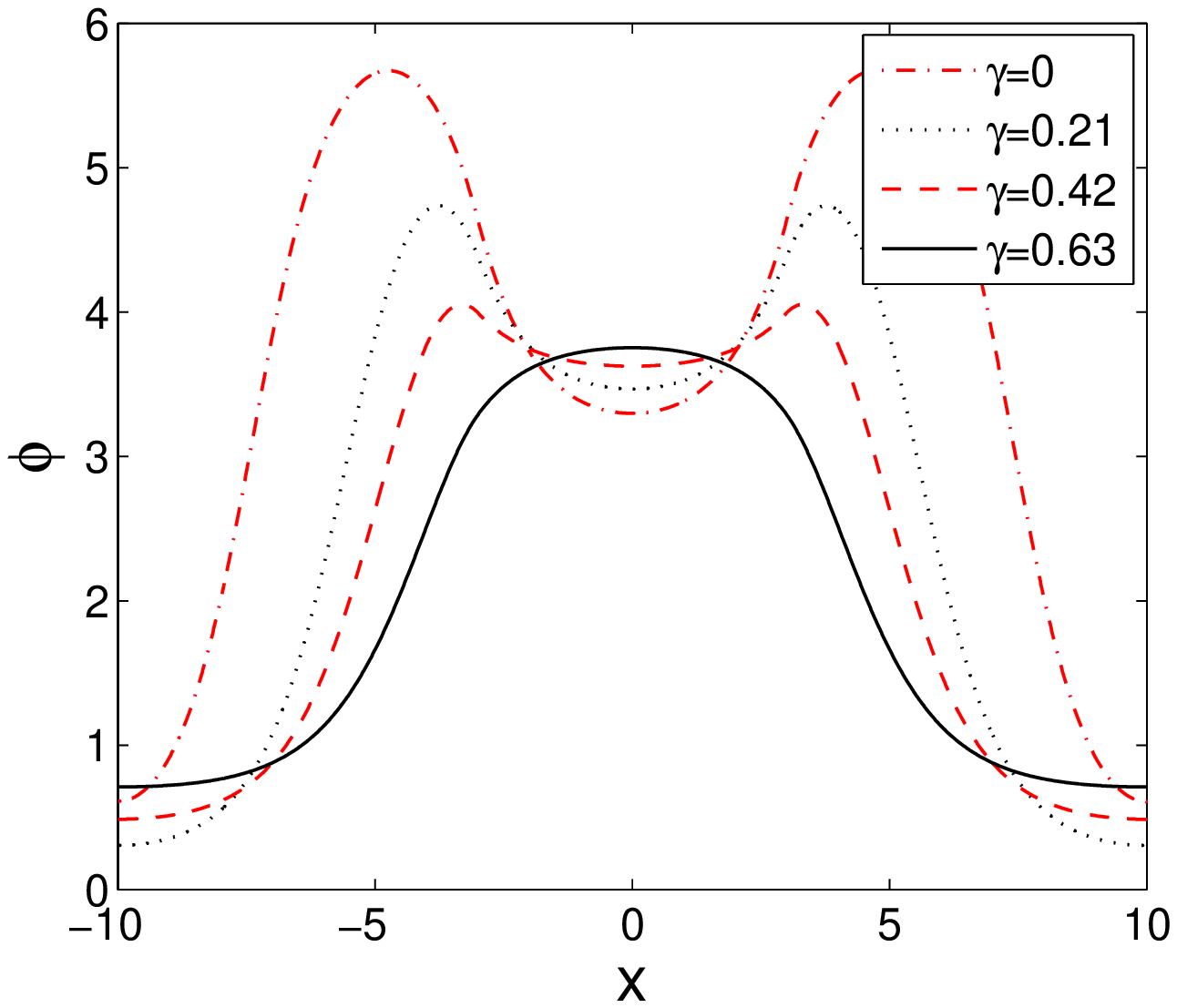}\\
\includegraphics[width=7cm]{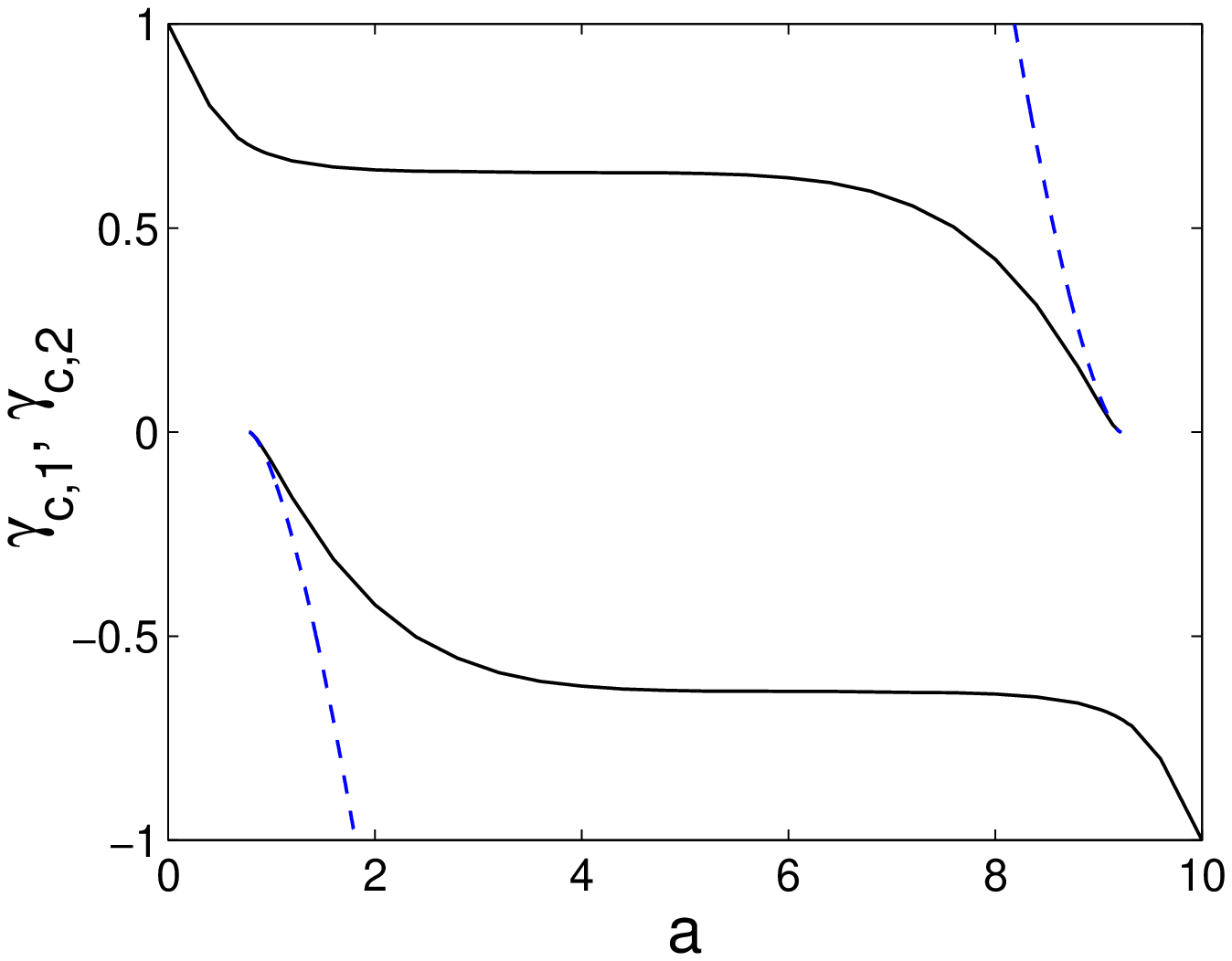}
\end{center}
\caption{(Color online) Top panel is the same as the top panel of Fig.\ \ref{gnonzero}, but for $L=10$ and $a=3$. Middle panel shows some of the corresponding solutions of the top branch for different values of $\gamma$. Bottom panel shows $\gamma_{c,1}$ and $\gamma_{c,2}$ as a function of $a$. Solid and dashed lines are numerical calculations and approximations (\ref{gappr}). }
\label{Llarge}
\end{figure}

Next, we study the influence of an applied bias current to the existence and the stability of 
the 
non constant ground state. In the following, we particularly consider $a=0.495$ and without loss of generality the 'positive' ground state, indicated by $\phi(0)>0$. The case of negative $\phi(0)$ can be obtained simply by 
 reflection due to the symmetry $\phi\to-\phi$ and $\gamma\to-\gamma$.

In the top panel of Fig.\ \ref{gnonzero}, we plot our numerical 
$\phi(0)$ as a function of the applied bias current $\gamma$ for $L=1$ and $a=0.495$.  
We use a path-following method starting from $\gamma=0$. 

First, we decrease the applied bias current. As $\gamma$ 
is reduced, the value of $\phi(0)$ 
also 
decreases up to a certain value of bias current; the solution cannot be continued further, 
it terminates in a saddle node bifurcation. Using our path following method, the saddle-node bifurcation is indeed due to a collision with a non-constant solution bifurcating from $\tilde{\phi}=0$, as 
predicted by our analytical result. The value of $\gamma$ at which the bifurcation occurs is 
the aforementioned $\gamma_{c,1}$ (\ref{gcappr}). Comparisons between the numerics and the analytical results of $\gamma_{c,1}$ are depicted at the middle panel of Fig.\ \ref{gnonzero}. 

Besides decreasing $\gamma$, one can also increase it. As $\gamma$ increases, the value of $\phi(0)$ 
also increases. As the bias current is increased further, 
a saddle-node bifurcation occurs. We denote this critical value of bias current by $\gamma_{c,2}$. For $a=0.495$, $|\gamma_{c,1}|<\gamma_{c,2}$. 
Using our path following algorithm, we 
can follow the upper branch of the bifurcation and deduce that it corresponds to a collision between the non constant solution and $\tilde{\phi}=\pi$. Using the cyclic symmetry argument, we 
explain the bifurcation using our analytical results (\ref{B0}). Plotted in the top panel of Fig.\ \ref{gnonzero} is our $\gamma(B_0)$ given by Eq.\ (\ref{gm}), properly shifted by $\pi$, for the facet length $(L-a)$. The influence of the $\pi$ facet length 
on the existence diagram is indicated by the arrows, i.e.\ as $a$ increases (decreases) towards $a_{c,\pi}$ the two lobes 
move according to the arrows (and vice versa). In the middle panel of Fig.\ \ref{gnonzero}, we 
 plot the second critical bias current, $\gamma_{c,2}$, as a function of the $\pi$ facet length $a$. Again, due to the cyclic symmetry, $\gamma_{c,2}$ can be obtained from $\gamma_{c,1}$ by rotating the curve by $\pi$ radians.

In the bottom panel of Fig.\ \ref{gnonzero}, we plot the critical eigenvalue of the non constant ground state as a function of $\gamma$ for $a=0.495$ and $L=1$. Interestingly, 
we observe that the lowest eigenvalue is attained at a non-zero bias current. This indicates that a non constant ground state can be made to be 'more stable' by applying 
a bias current. On the same figure, we also plot our approximations (\ref{EB0}), which 
qualitatively 
agree with the numerical results.

Studying further the saddle-node bifurcation between the non constant ground state and $\tilde\phi=\pi$ in Fig.\ \ref{gnonzero}, we observe that it is not the typical collision that leads to the definition of $\gamma_{c,2}$ for any $L$. When $L$ is relatively large, we 
find that the upper branch does not necessarily correspond to a constant solution. In Fig.\ \ref{Llarge}, we consider another case for $L=10$ and $a=3$.

Starting on the middle branch from $\gamma=0$ and $\tilde{\phi}(0)\approx 2.5$, we then increase the bias current. At the critical bias current $\gamma_{c,2}$, i.e.\ $\gamma\approx0.6$, 
we have a saddle-node bifurcation. Using our path following code, we follow the branch beyond the bifurcation point, from which we obtain that the branch does not correspond to 
a constant solution. In the middle panel of Fig.\ \ref{Llarge}, we plot the corresponding solutions for some values of $\gamma$. Considering the profile $\phi(x)$ at $\gamma=0$, we could conclude that it corresponds to a pair of semifluxons and one fluxon on each side. The profile is similar to the so-called semifluxon type 3, defined in Ref.\ \onlinecite{susa03} for an infinitely long 0-$\pi$ Josephson junction. From our numerical computations (not shown here), the two 
branches seem to be distinguished by the 
ability of the junction of length $L$ 
 to support 
  an additional fluxon on both sides. 
   The first critical value $\gamma_{c,1}$
   corresponds to the collision between the non constant ground state and $\phi=0$. The bottom panel presents $\gamma_{c,1}$ and $\gamma_{c,2}$ as a function of $a$ for $L=10$. An approximate expression 
   for the critical currents is presented 
    in dashed lines 
    given by\cite{ki,fn}
\begin{equation}
\begin{array}{lll}
&&\gamma_{c,1}= \displaystyle -\sqrt{\frac{128}{27(\pi+2)}}(a-a_{c,0})^{3/2},\\
&&\gamma_{c,2}= \displaystyle \sqrt{\frac{128}{27(\pi+2)}}(a_{c,\pi}-a)^{3/2},
\end{array}
\label{gappr}
\end{equation}
where $a_{c,0}\approx\pi/4$ and $a_{c,\pi}\approx(L-\pi/4)$.

One may ask 
about 
 using 0-$\pi$-0 Josephson junctions 
 to observe macroscopic quantum tunneling. To answer the question, the reader is addressed to Ref.\ \onlinecite{gold08}, 
  which considers quantum tunneling of a semifluxon in a finite 0-$\pi$ junction, where it was concluded that finite 0-$\pi$ junctions do not provide a good playground to build a qubit.\cite{gold08} Using the similarity between the currently  used Neumann boundary conditions and periodic boundary conditions (see Section \ref{symmetry}), one may conclude that the 
  system considered here is also not a promising one in which to observe macroscopic quantum tunneling, as finite 0-$\pi$-0 junctions considered here 
  are made of two finite 0-$\pi$ junctions.

\section{Conclusion}

We have investigated analytically and numerically 0-$\pi$-0 Josephson junctions on a finite domain. We have shown that there is  
an instability region for constant solutions in which semifluxons 
are spontaneously generated. Using an Euler-Lagrange approximation, it has been shown 
that the existence of static semifluxons
depends on the length of the junction, the facet length, and the applied bias current. In addition the critical eigenvalue of the semifluxons has been discussed. 
 Numerical simulations have been presented, accompanying our analytical results.

For future investigations, the two dimensional version of Josephson junctions with phase-shifts, 
 $\theta=\pi$ in some areas and 
 $\theta=0$  elsewhere, will be considered. The effects of boundary conditions 
 on the stability of non constant ground states in such a system will certainly be of interest.
 These are works in progress and will be reported in future publications.


\appendix

\section{Coefficient matrices}

The coefficient matrices $M_n,\,n=1,2,3,4,$ used to derive the 'continuous' and the discrete spectrum of $\phi=0,\,\pi$ are given by
\begin{widetext}

   \begin{eqnarray}
  M_1=\left[\begin{array}{rrrrrr}
    \begin{array}{cccccc}
      \cos(\hat{\alpha}(L-a)) &-\cos(\hat{\beta}a) & \sin(\hat{\beta}a) & 0 \\
      \hat{\alpha}\sin(\hat{\alpha}(L-a)) &  -\hat{\beta}\sin(\hat{\beta}a) & -\hat{\beta}\cos(\hat{\beta}a)& 0 \\
      0 &\cos(\hat{\beta}a) & \sin(\hat{\beta}a) & -\cos(\hat{\alpha}(a-L))  \\
      0 & -\hat{\beta}\sin(\hat{\beta}a) & \hat{\beta}\cos(\hat{\beta}a)  & \hat{\alpha}\sin(\hat{\alpha}(a-L)) \\
    \end{array}
    \end{array}\right],
\label{saud4}
\end{eqnarray}

   \begin{eqnarray}
M_2=  \left[\begin{array}{rrrrrr}
    \begin{array}{cccccc}
      \cosh(\hat{\gamma}(L-a)) & -\cos({\hat{\beta}a}) & \sin({\hat{\beta}a)} & 0 \\
      \hat{\gamma}\sinh{\hat{\gamma}(L-a)}& -\hat{\beta}\sin({\hat{\beta}a}) &-\hat{\beta}\cos({\hat{\beta}a})& 0 \\
      0 & \cos(\hat{\beta}a) & \sin(\hat{\beta}a) & \cosh({\hat{\gamma}(a-L)}) \\
      0 & -\hat{\beta}\sin(\hat{\beta}a) &  \hat{\beta}\sin(\hat{\beta}a)& \hat{\gamma}\sinh({\hat{\gamma}(a-L)})
    \end{array}
    \end{array}\right],
\label{saud5}
\end{eqnarray}

\begin{eqnarray}
M_3=  \left[\begin{array}{rrrrrr}
    \begin{array}{cccccc}
      \cos(\hat{\beta}(L-a))& -\cos(\hat{\alpha}a) & \sin(\hat{\alpha}a) & 0  \\
      \hat{\beta}\sin(\hat{\beta}(L-a)) & -\hat{\alpha}\sin({\hat{\alpha}a}) &-\hat{\alpha}\cos({\hat{\alpha}a})& 0 \\
      0 & \cos(\hat{\alpha}a) & \sin(\hat{\alpha}a) & -\cos(\hat{\beta}(a-L)) \\
      0 & -\hat{\alpha}\sin(\hat{\alpha}a) &   \hat{\beta}\sin(\hat{\beta}a)&-\hat{\beta}\sin(\hat{\beta}(a-L)) \\

    \end{array}
    \end{array}\right],
\label{saud7}
\end{eqnarray}

    \begin{eqnarray}
M_4=    \left[\begin{array}{rrrrrr}
    \begin{array}{cccccc}
      \cos(\hat{\beta}(L-a)) & -\cosh{\hat{\gamma}a} & \sinh({\hat{\gamma}a}) & 0 \\
      \hat{\beta}\sin(\hat{\beta}(L-a)) & \hat{\gamma}\sinh({\hat{\gamma}a}) &-\hat{\gamma}\cosh({\hat{\gamma}a}) & 0 \\
      0 & \cosh({\hat{\gamma}a}) & \sinh({\hat{\gamma}a}) & -\cos(\hat{\beta}(a-L)) \\
      0 & \hat{\gamma}\sinh({\hat{\gamma}a}) & \hat{\gamma}\cosh({\hat{\gamma}a}) & \hat{\beta}\sin{\hat{\beta}(a-L)}
\end{array}
\end{array}\right]. \end{eqnarray}

\end{widetext}


\begin{thebibliography}{99}

\bibitem {bula77} L.N. Bulaevskii, V.V. Kuzii, and A.A. Sobyanin,
Pis'ma Zh. Eksp. Teor. fiz. {\bf25}, 314 (1977) [JETP Lett. {\bf 25}, 290 (1977)].

\bibitem {bula78} L.N. Bulaevskii, V.V. Kuzii, A.A. Sobyanin, and P.N. Lebedev,
Solid State Comm. {\bf 25}, 1053 (1978).

\bibitem {vavr06} O. V\'avra, S. Ga\v{z}i, D. S. Golubovi\'c, I. V\'avra, J. D\'erer, J. Verbeeck, G. Van Tendeloo, and V. V. Moshchalkov,
Phys. Rev. B {\bf74}, 020502 (2006).

\bibitem {tsue00} C.C. Tsuei and J.R. Kirtley,
Rev. Mod. Phys. {\bf 72}, 969 (2000).

\bibitem {guma07} A. Gumann, C. Iniotakis, and N. Schopohl, 
    Appl. Phys. Lett. 91, 192502 (2007).

\bibitem {ryaz01} V.V. Ryazanov, V.A. Oboznov, A.Yu. Rusanov, A.V. Veretennikov, A.A. Golubov, and J. Aarts,
Phys. Rev Lett. {\bf 86}, 2427 (2001).

\bibitem {base99} J.J.A. Baselmans, A.F. Morpurgo, B.J. van Wees, and T.M. Klapwijk,
Nature {\bf 397}, 43 (1999).

\bibitem {gold04} E. Goldobin, A. Sterck, T. Gaber, D. Koelle, and R. Kleiner, 
Phys. Rev. Lett. {\bf92}, 057005 (2004).

\bibitem {hilg03} H. Hilgenkamp, Ariando, H. J. H. Smilde, D.H.A. Blank, G. Rijnders, H. Rogalla, J.R. Kirtley, and C.C. Tsuei,
Nature {\bf 422}, 50 (2003).

\bibitem {frol08} S.M. Frolov, M.J.A. Stoutimore, T.A. Crane, D.J. Van Harlingen, V.A. Oboznov, V.V. Ryazanov, A. Ruosi, C. Granata, M. Russo, Nature physics 4, 32-36 (2008).

\bibitem {ki} T. Kato \& M. Imada, J. Phys. Soc. Jpn.
{\bf 66}, 1445 (1997).

\bibitem {zenc04} A. Zenchuk and E. Goldobin, Phys. Rev. B 69, 024515 (2004).

\bibitem {susa03} H. Susanto, S.A. van Gils, T.P.P. Visser, Ariando, H.J.H. Smilde, and H. Hilgenkamp,
Phys. Rev. B 68, 104501 (2003).

\bibitem {gold04_2} E. Goldobin, D. Koelle, and R. Kleiner, Phys. Rev. B 70, 174519 (2004).

\bibitem {gold05_1} E. Goldobin, K. Vogel, O. Crasser, R. Walser, W. P. Schleich, D. Koelle, and R. Kleiner, Phys. Rev. B. 72, 054527 (2005).

\bibitem {gold05_2} E. Goldobin, H. Susanto, D. Koelle, R. Kleiner, and S. A. van Gils, Phys. Rev. B 71, 104518 (2005).

\bibitem {dewe08} A. Dewes, T. Gaber, D. Koelle, R. Kleiner, and E. Goldobin, Phys. Rev. Lett. 101, 247001 (2008).

\bibitem {hans06} J.A. Boschker, \emph{Manipulation and on-chip readout of fractional flux quanta}, Master thesis, University of Twente, 2006.

\bibitem {nick93} R.W.D.\ Nickalls, The Mathematical Gazette 77, 354–359 (1993).

\bibitem {gold08}  E. Goldobin, K. Vogel, W. P. Schleich, D. Koelle, and R. Kleiner, arXiv:0812.2394.

\bibitem {fn} There exists a typo in the expression of the critical current in Ref.\ \onlinecite{ki}.

\end{thebibliography}
\end{document}